# Do psychic cells generate consciousness?

Mototaka Suzuki[1,2] and Jaan Aru[3]

**Affiliation**

1. Brain Engineering Laboratory, Graduate School of Frontier Biosciences, The University of Osaka, Osaka, Japan
2. Department of Cognitive and Systems Neuroscience, Swammerdam Institute for Life Sciences, Faculty of Science, University of Amsterdam, Amsterdam, The Netherlands
3. Institute of Computer Science, University of Tartu, Tartu, Estonia.

**Abstract**

Technological advances in the past decades have begun to enable neuroscientists to address fundamental questions about consciousness in an unprecedented way. Here we review remarkable recent progress in our understanding of cellular-level mechanisms of conscious processing in the brain. Of particular interest are the cortical pyramidal neurons—or "psychic cells" called by Ramón y Cajal more than 100 years ago—which have an intriguing cellular mechanism that accounts for selective disruption of feedback signaling in the brain upon anesthetic-induced loss of consciousness. Importantly, a particular class of metabotropic receptors distributed over the dendrites of pyramidal cells are highlighted as the key cellular mechanism. After all, Cajal's instinct over a century ago may turn out to be correct—we may have just begun to understand whether and how psychic cells indeed generate and control our consciousness.

**Main text**

1. Introduction

In his book[2], *Texture of the nervous system of man and the vertebrates*, published in 1899, Ramón y Cajal called cortical pyramidal neurons psychic cells. He argued that "The *substratum* of such high activities [memory, thought, judgement and volition] is probably no other than the aggregate of association neurons of the cerebral cortex, i.e., those cells that establish the association between all sensory central areas" (p. 13). Despite the fact that the statement was made over a century ago, it is only in the recent few decades that significant steps have been made in our understanding of pyramidal neurons and their relevance to consciousness.

Before reviewing the studies of cortical pyramidal neurons, it is fair to state that evidence does exist which questions the major contribution of cerebral cortex to consciousness (reviewed in e.g., refs. [5,6]). For instance, decorticated cats and rats show indistinguishable behaviors (with minor exceptions) from animals with intact cortex[7,8]; surgical removal of significant portions of cerebral cortex does not deprive epileptic patients of consciousness[9]. However, there are multiple levels of conscious states, and at least some types of consciousness likely involves the rich information processing occurring in the cerebral cortex or the thalamo-cortical network[10-13]. For the sake of the main theme of this chapter being the cellular-mechanisms of conscious processing, we set the debate aside and highlight the remarkable recent studies that revealed what is happening at the cellular level when we lose or regain consciousness.

2. Cortical layer 5 pyramidal neurons

Cortical layer 5 pyramidal neurons (L5p) are the major output units of each cortical column or module[14] (Figure 1); they project their axons to local, nearby, and distant cortical and subcortical areas (such as the thalamus, the striatum, brainstem nuclei, and the spinal cord)[15-17]. They are the most active excitatory neurons in the cortical column[18]. They have unique morphology in that they spread their dendrites not only around the cell body in layer 5 and adjacent layers 4 and 6, but also heavily in the most distant superficial layer 1. This anatomical feature assures that these cells can receive and integrate all the information coming from cells across all six layers of the cortex. It is therefore no surprise that these cells play a central role in many brain (mal)functions—perception[19,20], prediction[21], learning[22-24], memory[25], dissociation[26,27], and behavior[28,29]. It is also no surprise that for a long time researchers have suggested that these cells might be related to consciousness. For instance, in 1994, Crick and Koch wrote: "Thus a very simplistic answer to the question "Which neurons fire in such a way that they correlate with awareness?" would be "The large pyramidal cells in layer 5 that fire in bursts and project outside the cortical system!" It would be marvelous if this were true…"[30]

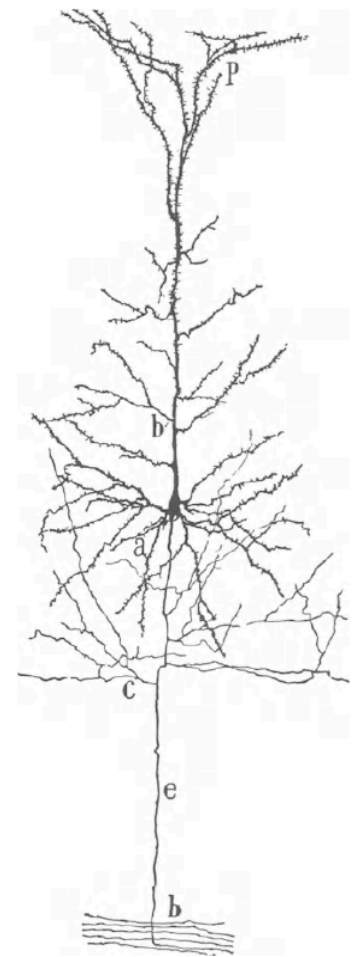

*Figure 1: Cajal's drawing of a pyramidal neuron of the rabbit cerebral cortex. a, basal dendrites; b, dendritic trunk and its branches; c, axon collaterals; e, long axon; p, distal apical dendrites. Adapted with permission from ref[2].*

3. Apical dendrites of cortical layer 5 pyramidal neurons

One unique feature of cortical layer 5 pyramidal neurons that endows the cells with unique capacities is their apical dendrites (Figure 1). L5p neurons spread their dendrites heavily in layer 1 where a high density of axonal terminals originates from cortical and subcortical areas[31,32]. Importantly, the information conveyed to layer 1 is predominantly feedback or top-down from higher-order cortical and subcortical areas[31-33]. Interestingly, the thalamic terminals are confined to the upper half of layer 1 (layer 1a)[34,35], whereas cortico-cortical projections are confined to the lower half of layer 1 (layer 1b)[24,34]. A conventional view of dendrites is that they are passive receivers of synaptic inputs from other cells, and the propagation of postsynaptic activity toward the soma depends on the distance from the synapses to the cell body (the cable theory[36]). According to this view, the inputs arriving at the very end of distal apical dendrites are the least effective in evoking somatic action potentials due to the largest distance. Does it mean that the information arriving in layer 1 is the least important?

On the contrary, evidence suggests that the information conveyed from higher-order cortical and thalamic areas to layer 1 is important[37,38] with so many types of information arriving from different brain regions[31,32,37]. If so, why does the information arrive at the least effective location (according to the cable theory)? Moreover, layer 1 has the highest density of excitatory synapses among all cortical layers[39,40], and therefore, a significant amount of plastic changes of synapses takes place in layer 1. If the distal apical dendrites are least effective in evoking somatic action potentials, why do so many plastic changes bother to occur in the distal dendrites despite the huge metabolic cost[41]?

Accumulating evidence, however, shows that dendrites are not mere passive cables but actively participate in the information flow[42]. Dendritic membranes have a rich repertoire of receptors and ion channels that are activated or inactivated by various factors, which can control whether the apical dendrites can or cannot influence somatic processing. As the distal apical dendrites receive top-down information, this control essentially comes down to determining how much prior information affects ongoing processing.

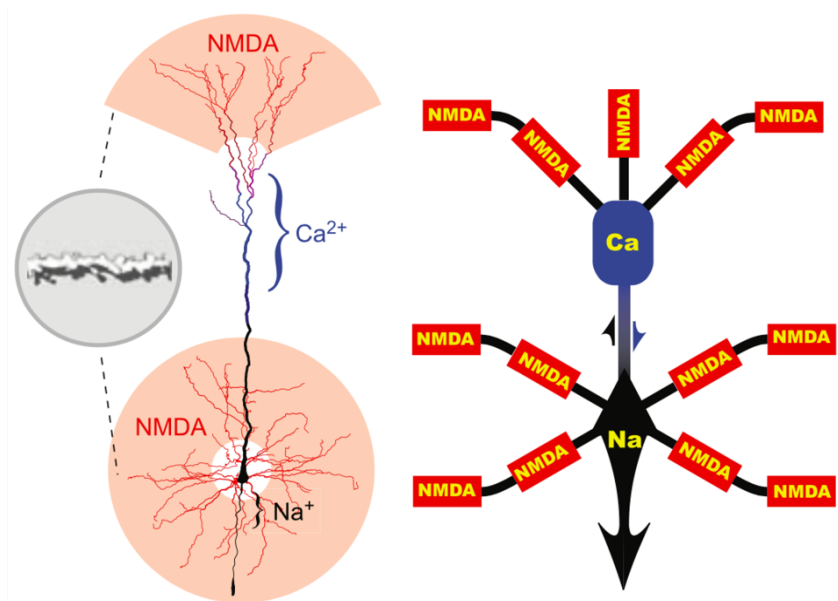

*Figure 2: Active properties of dendrites. Left, Reconstructed pyramidal neuron showing the regions of the dendritic tree where NMDA (red), Ca$^{2+}$ (blue), and Na$^+$ (black) electrogenesis can be initiated. (Inset) Enlargement of typical thin branch with a higher density of synapses than thicker dendrites[3]. Right, Schematic representation of the important subcompartments of a typical L5p neuron showing the relationship of multiple local sites for NMDA spikes to the Ca$^{2+}$ and Na$^+$ initiation zones, which can signal each other through active propagation along the main apical trunk (arrows). Adapted with permission from ref[4].*

For instance, one study revealed that inputs arriving at the distal apical dendrites of L5p can evoke regenerative calcium spikes by activating voltage-gated N-Methyl-D-aspartic acid (NMDA) receptors[4] (Figure 2, left); these spikes are called NMDA spikes. However, evoking a single NMDA spike is insufficient for evoking somatic action potentials reliably. L-type calcium channels—densely distributed around the primary bifurcation point of apical dendrites—can be activated by integrating NMDA spikes occurring at different dendritic branches, which induces another form of regenerative dendritic calcium event (Ca$^{2+}$ spike) that deterministically evokes a burst of somatic action potentials[4] (Figure 2, right). This active mechanism involving NMDA receptors and L-type calcium channels can reliably convey the inputs arriving at the distal end of apical dendrites to the cell bodies.

4. Dendro-somatic coupling and consciousness

Here let us define the dendro-somatic coupling as the degree to which the dendritic activity propagates toward the cell body. For example, if the dendritic activity propagates toward the cell body without attenuation and evokes somatic action potentials reliably, the dendro-somatic coupling is high; if the dendritic activity is instead attenuated along the way to the soma and evokes no somatic action potential, the dendro-somatic coupling is low or zero, i.e., decoupling.

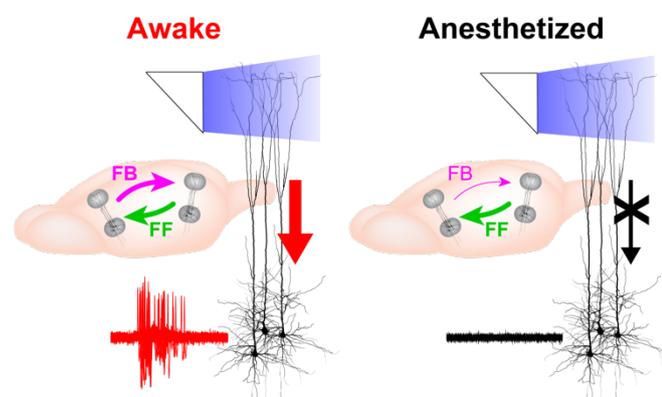

Our previous study addressed the question of whether the dendro-somatic coupling of L5p neurons changes across different conscious states[1] (Figure 3). To simulate the feedback inputs that activate the distal apical dendrites, we developed a novel micro-optical probe[43] that allowed us to optogenetically stimulate the distal apical dendrites specifically. By measuring the extracellular field potentials along the apical dendrites in awake and anesthetized states, we found that the propagation of dendritic activity from distal apical dendrites to soma is greatly altered by the conscious state of the animal[1]. In awake states, optogenetic activation of distal apical dendrites is highly effective in evoking action potentials rapidly, whereas during anesthesia the propagation of dendritic activity toward soma is strongly impaired and evokes no, if any, action potential. Three different anesthetics (isoflurane, urethane, ketamine/xylazine) had the same disruptive influence. These findings were surprising to us because numerous studies have shown that feedforward signals (e.g., sensory inputs) arriving at perisomatic regions of L5p or L2/3 neurons could reliably evoke somatic action potentials (e.g., ref[44]). Since inputs to the distal apical dendrites are predominantly feedback signals from higher cortical or subcortical regions[31-33], this mechanism explains why general anesthesia selectively disrupts feedback signaling in the brain while feedforward signaling and basic neuronal functions remain relatively unchanged[11].

*Figure 3: General anesthesia decouples cortical pyramidal neurons. In awake states, the dendritic depolarization rapidly propagates toward the soma and evokes somatic action potentials, whereas during anesthesia the dendritic activity is strongly suppressed along the apical dendrites and evokes no, if any, action potential. The feedback inputs activating the distal apical dendrites were simulated by the focal optogenetic stimulation. This cellular mechanism accounts for selective disruption of feedback signaling by general anesthesia. Adapted with permission from ref[1].*

Consistent with these findings, a more recent study has found that general anesthesia synchronizes spontaneous activity of L5p neurons globally, but during this anesthetic-induced synchronization apical dendrites of L5p neurons were desynchronized from their somas[45]. Strikingly, the change of L5p synchrony (and L5p dendro-somatic asynchrony) coincided with the loss and recovery of consciousness. These observations were made consistently across different anesthetics, and no other cell type showed such global synchronization nor dendro-somatic desynchronization upon anesthetic-induced loss of consciousness.

The next obvious question is, what is the cellular mechanism that differentiates the dendritic propagation in L5p neurons between awake and anesthetized states? Among many candidate mechanisms, we found that two types of metabotropic receptors—metabotropic glutamate receptors (mGluRs) and muscarinic acetylcholine receptors (mAChRs)—significantly contribute to the dendro-somatic coupling. Metabotropic receptors are G-protein coupled receptors and have multiple subtypes, e.g., mGluRs have eight subtypes, of which group I mGluRs (mGluR1 and mGluR5) are coupled with $G_q$ that induces the internal calcium release from endoplasmic reticulum[46]. This internal calcium increase might be the downstream mechanism that regulates the dendro-somatic coupling. There could also be interactions between mGluRs and mAChRs in L5p neurons like the synergistic effect found in hippocampal pyramidal neurons[47].

The dendro-somatic decoupling is likely happening during sleep-induced loss of consciousness as well. A previous study found that during sleep spindles[48,49]—characteristic brain waves that occur during non-rapid eye movement (NREM) sleep and are important for brain development and plasticity—the activity of apical dendrites of L5p neurons was synchronized with the spindles, but the activity of L5p somas was not[50]. Such a dendro-somatic desynchronization was not observed in layer 2/3 pyramidal neurons. Together with the findings during anesthesia, these studies suggest that the dendro-somatic decoupling is the common cellular mechanism during unconsciousness induced by two completely different methods—general anesthesia and natural sleep.

For a deeper understanding of consciousness, we need to be able to reversibly manipulate the conscious state, although it is undoubtedly valuable to study the brain in irreversible states of unconsciousness (e.g., coma, vegetative state, minimally conscious state)[51-53]. To this end, both general anesthesia and sleep—in particular, NREM sleep—continue to be beneficial for studying the cellular-level neuronal properties when losing and regaining consciousness. It would also be insightful to further characterize the intricate commonalities and differences as well as interactions between general anesthesia and sleep[54].

5. Psychic cells in context

So far we have only described the observations related to the cortex and cortical pyramidal neurons. However, even though L5p cells are the key integrators of cortical processing, their true effect becomes apparent when understood in the larger context of the thalamocortical system. Namely, The L5p cells are situated at the strategically central position to integrate corticocortical and thalamocortical processing. L5p cells are influenced by the higher-order (HO) thalamus, and they themselves also project to the HO thalamus. Hence, the L5p cells and their coupling state can control the large-scale dynamics of the brain.

Based on these cellular- and circuit-level facts we have proposed the Dendritic Integration Theory (DIT)[17,55]. DIT suggests that during conscious states, if the L5p cells are in the coupled state, activity can freely propagate activation patterns, activate the HO thalamus and the connected corticocortical loops, and give rise to global brain dynamics associated with consciousness. Such activity propagation cannot happen if the L5p cells are in a decoupled state. Hence, decoupling single L5p neurons decouples the thalamocortical loops and switches off the complex activity patterns underlying conscious processing[55,56]. In other words, this local switch happening at the level of single L5p neurons can change the dynamics of the whole thalamocortical system.

Previous studies have found that electrical stimulation of central thalamus partially restored the consciousness of patients who were in minimally conscious state[52] or anesthetized macaque monkeys[57-59]. Redinbaugh and colleagues found that the monkey's conscious level was correlated with the activity of the central thalamus and deep cortical neurons including L5p neurons[57]. This finding is consistent with our previous finding of L5p neurons' mechanism for selective suppression of feedback signaling[1]. Future studies may determine whether the central thalamus (and other higher-order thalamic nuclei) promotes consciousness by activating the dendro-somatic coupling of L5p neurons.

Although the crucial contribution of subcortical regions to consciousness is undeniable, it is still possible that their actions upon their projection target, i.e., the cortex, are critical. Indeed, central thalamic stimulation did enhance the conscious level of human patients or anesthetized monkeys, but only partial restoration of consciousness was possible[52,57], suggesting that thalamic stimulation alone is insufficient for full-blown consciousness. Interestingly, a recent study indeed found that pharmacological activation of cholinergic receptors in the medial prefrontal cortex accelerated emergence from anesthesia[60], which is consistent with our previous finding that activation of mAChRs is necessary for the dendro-somatic coupling in L5p neurons[1]. Together with the findings that other consciousness-associated higher-order thalamic[61,62] and brainstem nuclei[63,64] that all project widely to the cortex, it is tempting to speculate that the cortical mechanisms such as the dendro-somatic coupling can be one of the critical points of action exerted by all subcortical structures. Future studies may address this possibility.

6. Cellular mechanisms and theories of consciousness

We have proposed that the cellular mechanisms reviewed here could underlie many of the more global computations thought to be associated with consciousness[55,56]. For instance, the recurrent processing theory of consciousness (Lamme, this volume) and the predictive processing theory of consciousness (Pennartz, this volume) both rely on feedback connections being important. The present details about the cellular mechanisms directly show how the influence of feedback can be controlled in thalamocortical circuits.

Also, these cellular mechanisms are compatible with the global neuronal workspace theory of consciousness as dendro-somatic coupling could be the key mechanism for controlling ignition into the global workspace (Naccache, this volume). According to the integrated information theory of consciousness (IIT), consciousness corresponds to the neuronal processing complex with maximally irreducible cause–effect power on itself (Tononi, this volume). Dendro-somatic coupling could be the cellular mechanism for controlling information integration in the thalamocortical system. Modeling work has demonstrated that hindering dendritic integration leads to a drastic breakdown in the cause–effect repertoire[65].

7. Conclusions

We reviewed recent neurobiological findings suggesting that apical dendrites of L5p neurons play a central role in conscious processing in the brain. Since consciousness is the product of the brain, consciousness will eventually need to be explained with neurobiological terms. To this end, as Cajal called L5p neurons psychic cells, we believe that studying L5p neurons continues to be vital for our understanding of consciousness. Dendritic Integration Theory we derived from the findings in L5p neurons posits that L5p dendrites—more precisely, metabotropic receptors in the apical dendrites of L5p neurons—serve as the switch of consciousness by regulating the feedback signaling along the apical dendrites to the cell bodies[55]. Unlike many abstract theories of consciousness, this hypothesis is experimentally testable. We believe that this theory-driven study of L5p dendrites and their intradendritic mechanisms will be highly beneficial for consciousness research, even if this particular hypothesis eventually turns out to be incorrect. It would still contribute to the progress in our understanding of consciousness by deepening our knowledge about the neurobiological mechanisms underlying consciousness and whether Cajal's 'psychic cells' are part of the solution to the enigma.